\documentclass[preprint,12pt]{elsarticle}

\usepackage{epsfig}
\usepackage{enumerate}

\usepackage{amssymb,amsmath,graphics,graphicx}  
\usepackage[dvipsnames]{xcolor}

\journal{Elsevier}

\begin{document}

\begin{frontmatter}


\title{Modeling the impact of civilian firearm ownership in the evolution of violent crimes}

\author{Nuno Crokidakis}
\ead{nuno@mail.if.uff.br}

\address{Instituto de F\'{\i}sica, Universidade Federal Fluminense, Niter\'oi, Rio de Janeiro, Brazil}

\begin{abstract}
We propose a simple mathematical model to describe the evolution of violent crimes. For such purpose, we built a model based on ordinary differential equations that take into account the number of violent crimes and the number of legal and illegal guns. The dynamics is governed by probabilities, modeling for example the police action, the risk perception regarding crimes that leads to increase of ownership of legal guns, and so on. Our analytical and numerical results show that, in addition to the rise of criminality due to the presence of illegal guns, the increase of legal guns leads to a fast increase of violent crimes, suggesting that the access of firearms by civilians is not a good option regarding the control of crimes.

\end{abstract}


\end{frontmatter}

%\linenumbers


\section{Introduction}

All countries in the world face problems with criminality. The control of the number of committed violent crimes is one of the greatest problems of governments. The problem of criminality was studied by a series of researchers like mathematicians, physicists and computer scientists. For a recent review, see \cite{perc}.

In some countries, like United States, firearms are one of the central flashpoints in people's life \cite{buttrick}. The contemporary American desire to own a gun, but it is a desire with  consequences. Indeed, USA has more civilian-owned guns than any other developed nation and as a direct result has a level of gun-related deaths unmatched by any other developed nation \cite{buttrick}. Surveys revealed that about four-in-ten US adults live in a gun-owing househould. Thus, roughly $40\%$ of American households own a gun (about $30\%$ of American adults), and half of those Americans who do not currently own a gun can see themselves doing so in the future \cite{parker}. Thus, taking the population size of US as about 334 million \cite{worldometers}, $40\%$ means about 134 million guns. It was discussed that a record number of firearm background checks were completed in USA at the onset of the COVID-19 pandemic and during the protests following the murder of George Floyd \cite{lang}. Gun ownership is an expression of fear, but some analysis did suggest that people who own guns tend to exhibit lower levels of fear than nongun owners \cite{arrow}.

The usual questions are how and why criminals acquire, carry, and use guns. Key findings were that criminals rarely acquired firearms through conventional retail markets, and that criminals were for various reasons adept at finding ways to circumvent stricter gun controls \cite{wright}. The impact of not identifying the source of a weapon is vital to combating gun-enabled crime; failure to do so means that supply routes or ``armourers'' are not identified, allowing further supply of firearms into the criminal world \cite{elliott}.

Data from 13 European Union countries suggest that the number of illegally held firearms in the European Union is estimated to be up to 67 million. It is estimated that illicit trafficking has been directly responsible for at least 10,000-15,000 firearms-related deaths in European Union member states over the past decade \cite{duquet}. An analysis of the relationship between gun ownership rates and violent death rates in Europe indicates a strong positive correlation between gun ownership rates in a country and the rate of firearms-related deaths. \cite{duquet}. Regarding the civilian gun ownership in Europe, it was estimated in 93 million. The countries with more firearms in civilians' hands are Germany (25 million) and France (19 million) \cite{duquet}.

Unfortunately, research on the effect of gun levels on homicide and other crime rates has generally been of poor quality \cite{kleck}. Mathematical models can be useful to analyze some aspects of the evolution of crimes due to legal and/or illegal firearms. Indeed, we can cite some recent works \cite{monteiro,pritam,nuno,sooknanan,short}. In order to study specifically the impact of legal firearms on the evolution of violent crimes, i.e., crimes commited by civilians using legal guns, we propose an extension of the model presented in \cite{monteiro}. Our analytical and numerical results show that the increase of legal guns leads to a fast increase of violent crimes, suggesting that the access of firearms by civilians is not a good option regarding the control of crimes.

%%%%%%%%%%%%%%%%%%%%%%%%%%%

\section{The Model}

We consider the evolution of the number of violent crimes committed by an individual or a group of individuals. These crimes can be, for example, assaults, robberies and homicides, and the individual can be a gang member or a thief. These crimical activities are committed by using illegal guns, and it can be inhibited by the police action. In a recent work, the impact of illegal guns in the evolution of violent crimes was studied, and the author analyzed how legally-obtained guns can inhibit such crimes \cite{monteiro}. However, it was not considered the impact of such legally-obtained guns by citizens in the evolution of violent crimes, specifically in homicides. Recent studies analyzed the impact of homicides comitted by nonstragers, like husband, wife, mother, father, son, boyfriend, and others. For example, in USA, a parent was the perpetrator in the majority of homicides of children under age 5: $63\%$ were killed by a parent (1980 - 2008), $28\%$ by a friend/acquaintance and just $3\%$ by a stranger \cite{cooper}. In another work, it was observed that the average proportion of firearm homicides that were committed by strangers throughout the United States was only $21.9\%$ \cite{siegel2}. Motivated by those data, we also consider in the model crimes (firearm homicides) committed by citizens that own legal firearms.

Following the notation of Ref. \cite{monteiro}, let $x(t)$ be the prevalence of violent crimes per capita, $y(t)$ the number of legal firearms per capita and $z(t)$ the number of illegal firearms per capita. The following system of differential equations governs the dynamics of the model:
\begin{eqnarray} \label{eq1}
\frac{dx}{dt} & = & \alpha\,z - \nu\,x +\lambda\,y \\  \label{eq2}
\frac{dy}{dt} & = & \sigma + \beta\,x - \theta\,y - \gamma\,y\,z \\ \label{eq3}
\frac{dz}{dt} & = & \gamma\,y\,z + \delta\,z - \epsilon\,z^{2}
\end{eqnarray}

Let us elaborate on the model's parameters. Regarding Eq. \eqref{eq1}, the parameter $\alpha$ models the increase of violent crimes due to the presence of illegal guns. Typically we are talking about assaults, robberies and homicides. The negative term with the parameter $\nu$ represents the decrease of crimes due to the police's work. Finally, the parameter $\lambda$ is the novelty of the model, and it intends to take into account the rise of violent crimes due to legal firearms, as above discussed. We are considering that civilians only buy legal firearms ($y$). Thus, in this sense $y$ quantifies the civilian gun ownership, since more legal guns mean more guns in civilians' hands. In such a case, the increase of civilian gun ownership ($y$) leads to the increase of crimes ($x$), in addition to the crimes committed by illegal firearms ($z$). This justifies the positive term $+\lambda\,y$ in Eq. \eqref{eq1}. As a first approach to the problem of commitment of violent crimes by the use of legal guns by civilians, we considered the term as linear, but nonlinearities can also be considered in future versions of the model. In other words, the legal ownership of firearms by civilians leads to the commitment of crimes, for example a traffic fight can lead to a homicide if at least one of the civilians is a firearm owner, or the end of a marriage can lead the husband to kill his wife (or the wife to kill her husband) for not accepting the end of the relationship, if the husband/wife is a owner of a legal gun.

Looking now for Eq. \eqref{eq2}, the parameter $\sigma$ represents the input of legal firearms, that cannot depend on the present number of legal guns. The parameter $\beta$ models the legal guns' acquisition by the civil population, due mainly to the risk perception. In other words, if the percepection of violence is high, more guns are bought by civilians. Some governments have programs to control the guns' sales, like USA \cite{taylor}. Thus, this gun control by the State is modeled by the parameter $\theta$, that of course only act on the legal firearms. Indeed, fear of crime is correlated with attitude toward gun control \cite{hendrix}. All those terms are considered linear for simplicity. On the other hand, the last term in Eq. \eqref{eq2} with the parameter $\gamma$ models the situation in which a civilian with a legal gun's possession is robbed, and then that legal gun passes into the hands of an robber. In this way, this weapon is no longer a legal firearm because it changed its owner \cite{cook}. This term should be nonlinear since it models a kind of interaction among legal and illegal firearms.

Finally, we have Eq. \eqref{eq3}. The first term is related to the above-mentioned conversion of legal firearms into illegal ones. The term $\gamma\,y\,z$ enters in Eq. \eqref{eq2} as a negative term (decrease of legal firearms) and as a positive term in Eq. \eqref{eq3} (increase of illegal firearms). Finally, the two last terms with parameters $\delta$ and $\epsilon$ represent the increase of illegal guns. The increase of weapons cannot by unlimited, which justifies the two terms, one positive with parameter $\delta$ and other negative with $\epsilon$, that leads to the maximum value $\delta/\epsilon$ \cite{monteiro}. The terms can also be understood as follows. The positive term $+\delta\,z$ models in a simple way (linear function) how the current illegal firearms can be used to commitment of crimes such as assaults, robberies and homicides. These crimes lead the criminals to acquire money, that can be used to buy more illegal firearms. On the other hand, the negative term $-\epsilon\,z^{2}$ represents the confiscation of illegal firearms in police's actions, that is considered nonlinear in order to garantee a sustainable amount of illegal arms.

From the above Equations \eqref{eq1} - \eqref{eq3}, we took into account only $z$ presents a saturation, according to the logistic form considered in Eq. \eqref{eq3}. Of course also the time evolution of variables $x$ and $y$ can be modeled considering similar saturation terms. However, we are looking for a minimal model, i.e., the simplest form for the equations. In such a case, we are not considering bounds for $x$ and $y$, only for $z$. However, we will see in the following that the set of equations \eqref{eq1} - \eqref{eq3} leads to time evolutions that achieve equilibrium values for $x, y$ and $z$. Notice that the proposed dynamical system of Eqs.  \eqref{eq1} - \eqref{eq3} can be the starting point for the development of more complex models relating civilian gun ownership and the commitment of violent crimes.

For simplicity, we did not considered the fear of a criminal encountering an armed civilian. This kind of social interaction can be modeled by an extra term $-\mu\,y\,z$ in Eq. \eqref{eq1}, as investigated by Monteiro \cite{monteiro}. However, as discussed in \cite{kleck2}, in the case of USA, only about $1\%$ of the guns in private hands are used for defensive purposes. In this case, the mentioned parameter $\mu$ should be negligible.

%%%%%%%%%%%%%%%%%%%%%%%%%%%

\section{Results}

\begin{figure}[t]
\begin{center}
\vspace{6mm}
\includegraphics[width=0.7\textwidth,angle=0]{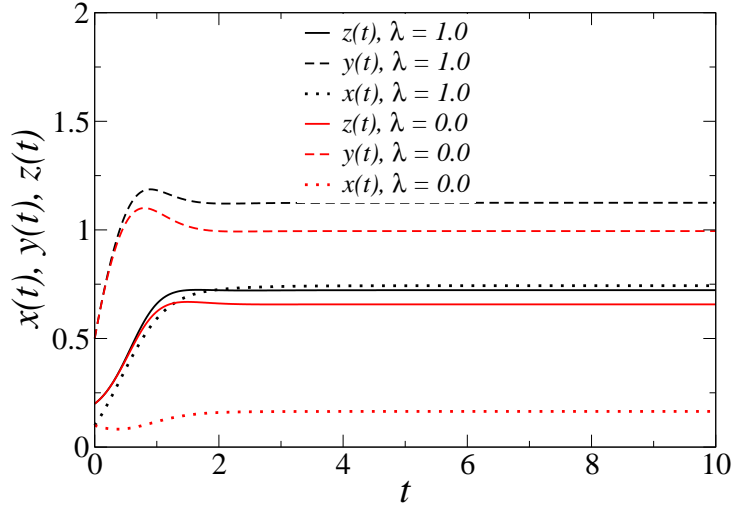}
\end{center}
\caption{(Color online) Illustration of the time evolution of the quantities $x(t)$ (dotted lines), $y(t)$ (dashed lines) and $z(t)$ (full lines), for the cases where crimes are committed by civilians that owner legal guns ($\lambda=1.0$, red curves) and where crimes are not committed by civilians, only by illegal guns ($\lambda=0.0$, black curves). The parameters are $\alpha=0.5, \beta=0.8, \gamma=2.5, \delta=0.8, \epsilon=5.0, \theta=0.5$ and $\nu=\sigma=2.0$ (all parameters are given in unities year$^{-1}$), and the initial conditions are given by $x(0)=0.1, y(0)=0.5$ and $z(0)=0.2$.}
\label{fig1}
\end{figure}

Let us start analyzing the time evolution of the previous defined quantities $x, y$ and $z$. We numerically integrated the Eqs. \eqref{eq1} - \eqref{eq3}, and the results are exhibited in Figure \ref{fig1}. The parameters are given in the caption of the figure. All parameters are rates, i.e., probabilities per unit time, and are given in unities year$^{-1}$. Specifically, as $\lambda$ is the new parameter of the model, we fixed all the other parameters and we exhibit the time evolution of $x, y$ and $z$ for two distinct values of $\lambda$, namely $\lambda=0.0$ and $\lambda=1.0$, in order to compare the model with crimes committed by legal guns ($\lambda=1.0$) with the model of Ref. \cite{monteiro}, where such crimes were not considered ($\lambda=0.0$) \footnote{For a detailed comparison of the models, we need to take $\lambda=0.0$ in our model and $\mu=0.0$ in the model of Ref. \cite{monteiro}.}. First of all, we can see that the quantities evolve in time and after a transient the system reaches stationary states with stable values of $x, y$ and $z$. These equilibrium values will be described analytically after an initial analysis of the time evolution. From Figure \ref{fig1} we can see that a nonzero value of $\lambda$ has a great impact in the evolution of violent crimes $x(t)$. First of all, we observe a fast increase of $x(t)$ for initial times for $\lambda=1.0$. On the other hand, for $\lambda=0.0$, $x(t)$ decreases for initial times and after it increases slowly and reaches the stationary state. Numerically, we found for $\lambda=0.0$ the stationary value $x\approx 0.1643$, whereas for $\lambda=1.0$ we have the equilibrium value $x\approx 0.7431$. In other words, the consideration of a relatively small value as $\lambda=1.0$ leads the equilibrium value of violent crimes per capita to increase about $4.5$ times in comparison with the absence of violent crimes committed by legal firearms \cite{monteiro}. This considerable difference is the first impact of $\lambda$ in the model, in comparison with Ref. \cite{monteiro}. The other quantities, $y$ and $z$, present a similar rate of increase for initial times, but for $\lambda=1.0$ the stationary values are greater than the case $\lambda=0.0$ \cite{monteiro}. Numerically, we found $(y,z)=(0.9945,0.6572)$ for $\lambda=0.0$ and $(y,z)=(1.1249,0.7224)$ for $\lambda=1.0$. 

%%%%%%%%%%%%%%%%%%%%%%%%%%%%%%%%%%%%%%%%%%%%%%%%%%%%%%%%%%%%%%%%
%%%%%%%%%%%%%%%%%%%%%%%%%%%%%%%%%%%%%%%%%%%%%%%%%%%%%%%%%%%%%%%%

As we observed in Figure \ref{fig1}, the quantities $x(t), y(t)$ and $z(t)$ evolve in time, and after some time they stabilize. In such stationary states, the time derivatives of Eqs. \eqref{eq1} - \eqref{eq3} are zero. In the $t\to\infty$ limit, Eq. \eqref{eq3} gives us two solutions, namely $z=0$ (extinction of illegal firearms) or
\begin{equation} \label{eq1n}
y = \frac{\epsilon\,z - \delta}{\gamma} ~.
\end{equation}  

If the solution $z=0$ is valid, Eq. \eqref{eq1} leads to
\begin{equation} \label{eq2n}
y=(\nu/\lambda)\,x ~. 
\end{equation}

Again considering the solution $z=0$ and take into account the result \eqref{eq2n}, Eq. \eqref{eq2} gives us
\begin{equation} \label{eq3n}
x = \frac{\lambda\,\sigma}{\theta\,\nu - \lambda\,\beta} ~.
\end{equation}

Substituting Eq. \eqref{eq3n} in \eqref{eq2n}, we found 
\begin{equation} \label{eq4n}
y = \frac{\nu\,\sigma}{\theta\,\nu - \lambda\,\beta} ~. 
\end{equation}
This set of solutions given by Eqs. \eqref{eq3n}, \eqref{eq4n} and $z=0$ represents the first equilibrium point $(x_{1}^{*}, y_{1}^{*}, z_{1}^{*})$.

The second equilibrium point $(x_{2}^{*}, y_{2}^{*}, z_{2}^{*})$ is obtained considering the solution $z\neq 0$. For such a case, we obtained above the result \eqref{eq1n} for $y$ such that $y=y(z)$. Taking the limit $t\to \infty$ in Eq. \eqref{eq1} and considering the result \eqref{eq1n}, we obtain a second solution for $x$, namely
\begin{equation} \label{eq5n}
x = \frac{1}{\nu}\left(\alpha+\frac{\lambda\,\epsilon}{\gamma}\right)z-\frac{\lambda\,\delta}{\gamma\,\nu} ~.
\end{equation}
In such a case, we have $x=x(z)$. Thus, Eqs. \eqref{eq5n} and \eqref{eq1n} give us the solutions for the equilibrium values of $x$ and $y$ as functions of $z$, for the case $z\neq 0$. This solution $z\neq 0$ can be obtained from Eq. \eqref{eq2} taking $t\to\infty$. We obtain $\sigma + \beta\,x = (\gamma\,z + \theta)\,y$. Substituing Eqs. \eqref{eq1n} and \eqref{eq5n} in this last result, we obtain a second order polynomial for $z$, namely $c_1\,z^{2} + c_2\,z + c_3 = 0$, where
\begin{eqnarray} \label{eq6}
c_1 & = & \gamma\,\nu\,\epsilon \\ \label{eq7}
c_2 & = & (\epsilon\,\theta - \gamma\,\delta)\,\nu - (\alpha\,\gamma+\lambda\,\epsilon)\,\beta \\ \label{eq8}
c_3 & = & \beta\,\lambda\,\delta - (\gamma\,\sigma + \delta\,\theta)\,\nu
\end{eqnarray} 

In other words, we have $z=\frac{-c_2 + \sqrt{c_2^{2}-4c_1c_3}}{2c_1}$, that can be written as
\begin{equation} \label{eq9}
z=\frac{1}{2\gamma\nu\epsilon}\{[(\alpha\gamma+\lambda\epsilon)\beta-(\epsilon\theta-\gamma\delta)\nu]+\sqrt{\Delta}\} ~,
\end{equation}
where
\begin{equation} \label{eq10}
\Delta=[(\alpha\gamma+\lambda\epsilon)\beta-(\epsilon\theta-\gamma\delta)]^{2}-4\gamma\nu\epsilon[\beta\lambda\delta-(\gamma\sigma+\delta\theta)\nu]  
\end{equation}

Eq. \eqref{eq9} completes the second equilibrium solution $(x_{2}^{*}, y_{2}^{*}, z_{2}^{*})$. Summarizing, the 2 equilibrium solutions are given by
\begin{eqnarray}\label{eq4}
(x_{1}^{*}, y_{1}^{*}, z_{1}^{*}) & = & \left(\frac{\lambda\,\sigma}{\theta\,\nu - \lambda\,\beta},\frac{\nu\,\sigma}{\theta\,\nu - \lambda\,\beta},0\right) \\ \label{eq5}
(x_{2}^{*}, y_{2}^{*}, z_{2}^{*}) & = & \left(\frac{1}{\nu}\left(\alpha+\frac{\lambda\,\epsilon}{\gamma}\right)z_{2}^{*}-\frac{\lambda\,\delta}{\gamma\,\nu},\frac{\epsilon\,z_{2}^{*}-\delta}{\gamma},z_{2}^{*}\right)
\end{eqnarray}

In other words, the presence of the parameter $\lambda$, that models the increase of violent crimes due to the presence of legal firearms in the civilians' hands, eliminates the crime-free equilibrium found in \cite{monteiro}. For our solution 1, we observe that the equilibrium number of illegal firearms per capita is zero ($z_{1}^{*}=0$) but the other quantities, namely the prevalence of violent crimes per capita $x_{1}^{*}$ and the number of legal firearms per capita $y_{1}^{*}$, are different from zero. Notice that if we take $\lambda=0$ in solution $(x_{1}^{*}, y_{1}^{*}, z_{1}^{*})$ we recover the result $(0, \frac{\sigma}{\theta},0)$, i.e., the crime-free equilibrium (since $x_{1}^{*}=0$) analyzed in \cite{monteiro}.

A simple stability analysis shows that the solution $(x_{1}^{*}, y_{1}^{*}, z_{1}^{*})$ is unstable, whereas the another one $(x_{2}^{*}, y_{2}^{*}, z_{2}^{*})$ is stable. Thus, let us analyze such solution 2. We observe that $x_{2}^{*}$ grows proportional to $z_{2}^{*}$, as in \cite{monteiro}, but the parameter $\lambda$ amplifies such effect. Thus, more illegal guns lead to more crimes, as observed in \cite{monteiro}. We can also see that as $x_{2}^{*}$ grows with $\lambda$, i.e., the number of violent crimes increases due to the presence of legal firearms in the civilians' hands, since more crimes can be committed by such legal guns (motivated by conflicting opinions or prejudice, among other, as discussed above). This is a realistic result, and it was not observed in the model of Ref. \cite{monteiro} since the author in \cite{monteiro} did not considered crimes committed by legal guns. The police action also decreases the number of crimes, since $x_{2}^{*}\propto 1/\nu$. In other words, if the state wants to fight homicides, evidences show that removing criminals from circulation with incarceration is the best option \cite{odon}.

%%%%%%%%%%%%%%%%%%%%%%%%%%%%%%%%%%%%%%%%%%%%%%%%%%%%%%%%%%%%%%%%
\begin{figure}[t]
\begin{center}
\vspace{6mm}
\includegraphics[width=0.7\textwidth,angle=0]{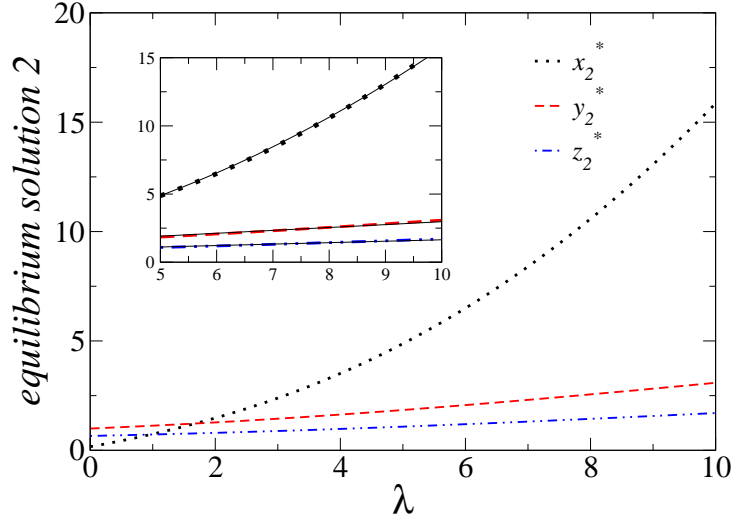}
\end{center}
\caption{(Color online) Equilibrium values  $x_{2}^{*}$ (dotted line), $y_{2}^{*}$ (dashed line) and $z_{2}^{*})$ (dotted-dashed line) as functions of $\lambda$. The lines are given by Eqs. \eqref{eq5} and \eqref{eq9}. The parameters are $\alpha=0.5, \beta=0.8, \gamma=2.5, \delta=0.8, \epsilon=5.0, \theta=0.5$ and $\nu=\sigma=2.0$ (all parameters are given in unities year$^{-1}$). In the inset we show a zoom in the region of large $\lambda$, and the full (black) lines exhibit the behaviors $x_{2}^{*}\propto \lambda^{2}, y_{2}^{*}\propto \lambda$ and $z_{2}^{*}\propto \lambda$, as discussed in the text.}
\label{fig2}
\end{figure}
%%%%%%%%%%%%%%%%%%%%%%%%%%%%%%%%%%%%%%%%%%%%%%%%%%%%%%%%%%%%%%%%

Regarding the mentioned dependence of the equilibrium solution $(x_{2}^{*}, y_{2}^{*}, z_{2}^{*})$ on the new parameter $\lambda$, a simple inspection of Eqs. \eqref{eq5} and \eqref{eq9} shows that for large $\lambda$ we have the following dependencies: $z_{2}^{*}\propto \lambda, y_{2}^{*}\propto \lambda$ and $x_{2}^{*}\propto \lambda^{2}$. Since $\lambda$ is the rate of increase of violent crimes ($x$) due to the presence of legal guns in the civilian' hands ($y$), the last results show that the civilian gun ownership leads to a fast increase of violent crimes, despite the slower growth of legal and illegal firearms. An illustration is given in Figure \ref{fig2}. We can see the fast growth of violent crimes for increasing $\lambda$, and slower growth of both legal and illegal firearms. Those results are in agreement with the previous discussions. The inset in Figure \ref{fig2} shows the same graphic for the larger values of $\lambda$, exhibiting the mentioned behaviors $z_{2}^{*}\propto \lambda, y_{2}^{*}\propto \lambda$ and $x_{2}^{*}\propto \lambda^{2}$. We also plot in the inset for comparison curves of the type $f(x)=ax^{2} + bx +c$ (for $x_{2}^{*}$) and $g(x)=ax+b$ (for $y_{2}^{*}$ and $z_{2}^{*}$), to better visualization of the mentioned quadratic and linear behaviors of the quantities of interest (see the full lines).

We can also discuss about the control of firearms by government. For this purpose, we can follow the strategy of Ref. \cite{monteiro} and consider two situations:
\begin{itemize}

\item (a) a first situation where the government tightly controls arms' sales. For such a case, we should take the following limits for the model's parameters: $\theta\to\infty$, $\sigma\to 0$, $\beta\to 0$

\item (b) a second situation where there is no strong control of guns.  For such a case, we should take the following limits for the model's parameters: $\theta\to 0$, $\beta\to\infty$

\end{itemize}

Considering situation (a), the equilibrium point 2, Eq. \eqref{eq5}, can be approximated to 
\begin{equation} \label{eq11}
(x_{2}^{*}, y_{2}^{*}, z_{2}^{*}) \approx \left(\frac{\alpha\delta}{\nu\epsilon},0,\frac{\delta}{\epsilon}\right)
\end{equation}
that is the same result of Ref. \cite{monteiro}. It means that if the government really control the sale of firearms, the legal guns can disappear of the population after a long time, and they cannot contribute to violent crimes. If the civilians do not own legal guns, the number of homicides can effectively decreases. Considering such strategy (a), the violent crimes can be reduced if the police action is increased in order to seize weapons. This is a difficult point, of course. For example, firearms registrations increased $120\%$ in 2020 in Brazil, but seizures are down \cite{brazil}.

On the other hand, for the situation (b) we have:
\begin{equation} \label{eq12}
(x_{2}^{*}, y_{2}^{*}, z_{2}^{*}) \approx \left(\frac{1}{\gamma\nu}\left[\frac{F^{2}\beta}{\gamma\nu\epsilon}-\lambda\delta\right],\frac{1}{\gamma^{2}\nu}[F\beta-\delta\gamma\nu],\frac{F\beta}{\gamma\nu\epsilon}\right)
\end{equation}
where
\begin{equation}
F = \alpha\gamma+\lambda\epsilon
\end{equation}

For such second situation (b), one can see that the increase of police action $\nu$ leads to the decrease of crimes $x_{2}^{*}$, as well as the decrease of illegal firearms $z_{2}^{*}$. For rising $\lambda$, crimes, legal and illegal guns increase, since the control of guns is weak. We also observe the same behavior as before, i.e., $z_{2}^{*}\propto \lambda, y_{2}^{*}\propto \lambda$ and $x_{2}^{*}\propto \lambda^{2}$. In other words, the number of legal firearms increases with $\beta$ and $\lambda$, which leads to $x_{2}^{*}\neq 0$. Comparing with Ref. \cite{monteiro}, the author found $x_{2}^{*}=0$. For our model, we see that the violent criminality cannot be eradicated in both situations (a) and (b). Since even in this limit of weak gun control the quantities $x, y$ and $z$ depend on $\lambda$, it is clear that a substantial number of per capita crimes will be committed by civilians if they own legal guns. This can be an important result regarding firearms' control \cite{duquet,kleck}.

%%%%%%%%%%%%%%%%%%%%%%%%%%%%%%%%%%%%%%%%%

\section{Final Remarks}   

In this work we study a simple model of violent crimes' evolution. For this purpose, we built a system of three ordinary differential equations for the time evolution of three quantities of interest: violent crimes $x$, legal firearms $y$ and illegal firearms $z$. Probabilities govern the evolution of such equations.

The model deals with the key point of crimes committed by citizens that own legal guns. As discussed in \cite{waiselfisz}, firearms in the hands of the population would increase the risk of any conflict or dispute ending in murder. Thus, the mathematical modeling of crimes committed by civilian population by the use of legal firearms is an important subject to be studied.

We found the following dependencies for the equilibrium point: $z_{2}^{*}\propto \lambda, y_{2}^{*}\propto \lambda$ and $x_{2}^{*}\propto \lambda^{2}$. As we considered that civilians only buy legal firearms, $y_{2}^{*}$ represents the equilibrium point for the number of legal firearms per capita, i.e., it quantifies the civilian gun ownership. Thus, since $\lambda$ represents the rate of increase of violent crimes due to the presence of legal guns, our results show that the increase of civilian gun ownership leads to a fast increase of violent crimes, despite the slower grow of legal and illegal firearms. Thus, more guns do not imply in less crimes, but the opposite occurs. Indeed, regarding firearm incidents in American schools, as observed by the author in \cite{hamlin}, State gun ownership rates declined between 1980 and 2019 while school firearm incidents generally ranged between 20 and 40 incidents before skyrocketing to 102 incidents in 2018 and 110 incidents in 2019. In addition, another recent study examinated the association between gun ownership rates and homicide rates in the United States from 1973 to 2016. The results show a very strong positive correlation between gun ownership and homicide rates, which supports the hypothesis that higher rates of gun ownership increase the likelihood of homicide \cite{smith}. The authors in \cite{smith} found that, on average, a $1\%$ increase in gun ownership is found to be associated with $2.6$ additional homicides per million people. In agreement with \cite{smith}, the authors in \cite{siegel} found that gun ownership was a significant predictor of firearm homicide rates: the results indicated that for each percentage point increase in gun ownership, the firearm homicide rate increased by $0.9\%$.

Our results show that a substantial number of per capita crimes will be committed by civilians if they own legal guns. The model shows that the violent criminality cannot be eradicated even considering strong and weak controls of guns. This can turn the violence more complicated in countries like Brazil, where the control of firearms adopted has not been shown to be effective in fighting crime or in reducing violence, specially in the number of homicides \cite{moura}. Thus, police action as well as stronger gun control is mandatory to control the occurrence of violent crimes.

%%%%%%%%%%%%%%%%%%%%%%%%%%%%%%%%%%%%%%%%%

\section*{Acknowledgments}

The author acknowledges financial support from the Brazilian scientific funding agencies Conselho Nacional de Desenvolvimento Cient\'ifico e Tecnol\'ogico (CNPq, Grant 310893/2020-8) and Funda\c{c}\~ao de Amparo \`a Pesquisa do Estado do Rio de Janeiro (FAPERJ, Grant 203.217/2017).

\bibliographystyle{elsarticle-num-names}

\end{document}